\documentclass[12pt]{article}
\usepackage{amssymb}
\usepackage{amsmath}

\begin{document}

\begin{titlepage}
\title{\bf Hamilton Dynamics on Clifford K\"{a}hler Manifolds}
\author{ Mehmet Tekkoyun \footnote{Corresponding author. E-mail address: tekkoyun@pau.edu.tr; Tel: +902582953616; Fax: +902582953593}\\
{\small Department of Mathematics, Pamukkale University,}\\
{\small 20070 Denizli, Turkey}}
\date{\today}
\maketitle

\begin{abstract}

This paper presents Hamilton dynamics on Clifford  K\"{a}hler
manifolds. In the end, the some results related to Clifford
K\"{a}hler dynamical systems are also discussed.

{\bf Keywords:} Clifford K\"{a}hler Geometry, Hamiltonian
Dynamics.

{\bf PACS:} 02.40.

\end{abstract}
\end{titlepage}

\section{Introduction}

Modern differential geometry explains explicitly the dynamics of Hamiltons.
So, if $Q$ is an $m$-dimensional configuration manifold and $\mathbf{H}%
:T^{\ast }Q\rightarrow \mathbf{R}$\textbf{\ }is a regular Hamilton function,
then there is a unique vector field $X$ on $T^{\ast }Q$ such that dynamic
equations are determined by
\begin{equation}
\,\,i_{X}\Phi =d\mathbf{H}  \label{1.1}
\end{equation}%
where $\Phi $ indicates the symplectic form. The triple $(T^{\ast }Q,\Phi
,X) $ is called \textit{Hamilton system }on the cotangent bundle $T^{\ast
}Q. $

At last time, there are many studies and books about Hamilton mechanics,
formalisms, systems and equations such that real, complex, paracomplex and
other analogues \cite{deleon, tekkoyun} and there in. Therefore it is
possible to obtain different analogous in different spaces.

It is known that quaternions were invented by Sir William Rowan Hamilton as
an extension to the complex numbers. Hamilton's defining relation is most
succinctly written as:

\begin{equation}
i^{2}=j^{2}=k^{2}=ijk=-1  \label{1.2}
\end{equation}%
If it is compared to the calculus of vectors, quaternions have slipped into
the realm of obscurity. They do however still find use in the computation of
rotations. A lot of physical laws in classical, relativistic, and quantum
mechanics can be written pleasantly by means of quaternions. Some physicists
hope they will find deeper understanding of the universe by restating basic
principles in terms of quaternion algebra. It is well-known that quaternions
are useful for representing rotations in both quantum and classical
mechanics \cite{dan} . It is well known that Clifford manifold is a
quaternion manifold. So, all properties defined on quaternion manifold of
dimension $8n$ also is valid for Clifford manifold. Hence, it may be
constructed mechanical equations on Clifford K\"{a}hler manifold.

The paper is structured as follows. In second 2, we review Clifford K\"{a}%
hler manifolds. In second 3 we introduce Hamilton equations for dynamical
systems on Clifford K\"{a}hler manifold. In conclusion, we discuss some
geometric-physical results about Hamilton equations and fields constructed
on the base manifold.

\section{Preliminaries}

In this paper, all mappings and manifolds are assumed to be smooth, i.e.
infinitely differentiable and the sum is taken over repeated indices. By $%
\mathcal{F}(M)$, $\chi (M)$ and $\Lambda ^{1}(M)$ we understand the set of
functions on $M$, the set of vector fields on $M$ and the set of 1-forms on $%
M$, respectively.

\subsection{Clifford K\"{a}hler Manifolds}

Here, we recall and extend the main concepts and structures given in \cite%
{yano, burdujan, tekkoyun1} . Let $M$ be a real smooth manifold of dimension
$m.$ Suppose that there is a 6-dimensional vector bundle $V$ consisting of $%
F_{i}(i=\overline{1,6})$ tensors of type (1,1) over $M.$ Such a local basis $%
\{F_{1},F_{2},...,F_{6}\}$ is named a canonical local basis of the bundle $V$
in a neighborhood $U$ of $M$. Then $V$ is called an almost Clifford
structure in $M$. The pair $(M,V)$ is named an almost Clifford manifold with
$V$. Thus, an almost Clifford manifold $M$ is of dimension $m=8n.$ If there
exists on $(M,V)$ a global basis $\{F_{1},F_{2},...,F_{6}\},$ then $(M,V)$
is called an almost Clifford manifold; the basis $\{F_{1},F_{2},...,F_{6}\}$
is said to be a global basis for $V$.

An almost Clifford connection on the almost Clifford manifold $(M,V)$ is a
linear connection $\nabla $ on $M$ which preserves by parallel transport the
vector bundle $V$. This means that if $\Phi $ is a cross-section
(local-global) of the bundle $V$, then $\nabla _{X}\Phi $ is also a
cross-section (local-global, respectively) of $V$, $X$ being an arbitrary
vector field of $M$.

If for any canonical basis $\{J_{i}\}$, $i=\overline{1,6}$of $V$ in a
coordinate neighborhood $U$, the identities
\begin{equation}
g(J_{i}X,J_{i}Y)=g(X,Y),\text{ }\forall X,Y\in \chi (M),\text{ }\
i=1,2,...,6,  \label{2.2}
\end{equation}%
hold, the triple $(M,g,V)$ is called an almost Clifford Hermitian manifold
or metric Clifford manifold denoting by $V$ an almost Clifford structure $V$
and by $g$ a Riemannian metric and by $(g,V)$ an almost Clifford metric
structure$.$

Since each $J_{i}(i=1,2,...,6)$ is almost Hermitian structure\ with respect
to $g$, setting

\begin{equation}
\Phi _{i}(X,Y)=g(J_{i}X,Y),~\text{ }i=1,2,...,6,  \label{2.3}
\end{equation}

for any vector fields $X$ and $Y$, we see that $\Phi _{i}$ are 6 local
2-forms.

If the Levi-Civita connection $\nabla =\nabla ^{g}$ on $(M,g,V)$ preserves
the vector bundle $V$ by parallel transport, then $(M,g,V)$ is named a
Clifford K\"{a}hler manifold, and an almost Clifford structure $\Phi _{i}$
of $M$ is said to be a Clifford K\"{a}hler structure. Suppose that let
\begin{equation*}
\left\{
x_{i},x_{n+i},x_{2n+i},x_{3n+i},x_{4n+i},x_{5n+i},x_{6n+i},x_{7n+i}\right\}
,i=\overline{1,n}
\end{equation*}%
be a real coordinate system on $(M,V).$ Then we denote by
\begin{eqnarray}
&&\left\{ \frac{\partial }{\partial x_{i}},\frac{\partial }{\partial x_{n+i}}%
,\frac{\partial }{\partial x_{2n+i}},\frac{\partial }{\partial x_{3n+i}},%
\frac{\partial }{\partial x_{4n+i}},\frac{\partial }{\partial x_{5n+i}},%
\frac{\partial }{\partial x_{6n+i}},\frac{\partial }{\partial x_{7n+i}}%
\right\} ,  \label{2.4} \\
&&%
\{dx_{i},dx_{n+i},dx_{2n+i},dx_{3n+i},dx_{4n+i},dx_{5n+i},dx_{6n+i},dx_{7n+i}\}
\notag
\end{eqnarray}%
the natural bases over $\mathbf{R}$ of the tangent space $T(M)$ and the
cotangent space $T^{\ast }(M)$ of $M,$ respectively$.$ By structures $%
\left\{ J_{1},J_{2},J_{3},J_{4},J_{5},J_{6}\right\} $ the following
expressions are given%
\begin{equation}
\begin{array}{ccc}
\begin{array}{c}
J_{1}(\frac{\partial }{\partial x_{i}})=\frac{\partial }{\partial x_{n+i}}
\\
\text{ }J_{1}(\frac{\partial }{\partial x_{n+i}})=-\frac{\partial }{\partial
x_{i}} \\
J_{1}(\frac{\partial }{\partial x_{2n+i}})=\frac{\partial }{\partial x_{4n+i}%
} \\
J_{1}(\frac{\partial }{\partial x_{3n+i}})=\frac{\partial }{\partial x_{5n+i}%
} \\
J_{1}(\frac{\partial }{\partial x_{4n+i}})=-\frac{\partial }{\partial
x_{2n+i}} \\
J_{1}(\frac{\partial }{\partial x_{5n+i}})=-\frac{\partial }{\partial
x_{3n+i}} \\
\text{ }J_{1}(\frac{\partial }{\partial x_{6n+i}})=\frac{\partial }{\partial
x_{7n+i}} \\
J_{1}(\frac{\partial }{\partial x_{7n+i}})=-\frac{\partial }{\partial
x_{6n+i}}%
\end{array}
&
\begin{array}{c}
J_{2}(\frac{\partial }{\partial x_{i}})=\frac{\partial }{\partial x_{2n+i}}
\\
J_{2}(\frac{\partial }{\partial x_{n+i}})=-\frac{\partial }{\partial x_{4n+i}%
} \\
J_{2}(\frac{\partial }{\partial x_{2n+i}})=-\frac{\partial }{\partial x_{i}}
\\
J_{2}(\frac{\partial }{\partial x_{3n+i}})=\frac{\partial }{\partial x_{6n+i}%
} \\
J_{2}(\frac{\partial }{\partial x_{4n+i}})=\frac{\partial }{\partial x_{n+i}}
\\
J_{2}(\frac{\partial }{\partial x_{5n+i}})=-\frac{\partial }{\partial
x_{7n+i}} \\
J_{2}(\frac{\partial }{\partial x_{6n+i}})=-\frac{\partial }{\partial
x_{3n+i}} \\
J_{2}(\frac{\partial }{\partial x_{7n+i}})=\frac{\partial }{\partial x_{5n+i}%
}%
\end{array}
&
\begin{array}{c}
J_{3}(\frac{\partial }{\partial x_{i}})=\frac{\partial }{\partial x_{3n+i}}
\\
J_{3}(\frac{\partial }{\partial x_{n+i}})=-\frac{\partial }{\partial x_{5n+i}%
} \\
J_{3}(\frac{\partial }{\partial x_{2n+i}})=-\frac{\partial }{\partial
x_{6n+i}} \\
J_{3}(\frac{\partial }{\partial x_{3n+i}})=-\frac{\partial }{\partial x_{i}}
\\
J_{3}(\frac{\partial }{\partial x_{4n+i}})=\frac{\partial }{\partial x_{7n+i}%
} \\
J_{3}(\frac{\partial }{\partial x_{5n+i}})=\frac{\partial }{\partial x_{n+i}}
\\
J_{3}(\frac{\partial }{\partial x_{6n+i}})=\frac{\partial }{\partial x_{2n+i}%
} \\
J_{3}(\frac{\partial }{\partial x_{7n+i}})=-\frac{\partial }{\partial
x_{4n+i}}%
\end{array}
\\
\begin{array}{c}
J_{4}(\frac{\partial }{\partial x_{i}})=\frac{\partial }{\partial x_{4n+i}}
\\
\text{ }J_{4}(\frac{\partial }{\partial x_{n+i}})=-\frac{\partial }{\partial
x_{2n+i}} \\
J_{4}(\frac{\partial }{\partial x_{2n+i}})=\frac{\partial }{\partial x_{n+i}}
\\
J_{4}(\frac{\partial }{\partial x_{3n+i}})=-\frac{\partial }{\partial
x_{7n+i}} \\
J_{4}(\frac{\partial }{\partial x_{4n+i}})=-\frac{\partial }{\partial x_{i}}
\\
\text{ }J_{4}(\frac{\partial }{\partial x_{5n+i}})=\frac{\partial }{\partial
x_{6n+i}} \\
J_{4}(\frac{\partial }{\partial x_{6n+i}})=-\frac{\partial }{\partial
x_{5n+i}} \\
J_{4}(\frac{\partial }{\partial x_{7n+i}})=\frac{\partial }{\partial x_{3n+i}%
}%
\end{array}
&
\begin{array}{c}
J_{5}(\frac{\partial }{\partial x_{i}})=\frac{\partial }{\partial x_{5n+i}}
\\
\text{ }J_{5}(\frac{\partial }{\partial x_{n+i}})=-\frac{\partial }{\partial
x_{3n+i}} \\
J_{5}(\frac{\partial }{\partial x_{2n+i}})=-\frac{\partial }{\partial
x_{7n+i}} \\
J_{5}(\frac{\partial }{\partial x_{3n+i}})=\frac{\partial }{\partial x_{n+i}}
\\
J_{5}(\frac{\partial }{\partial x_{4n+i}})=\frac{\partial }{\partial x_{6n+i}%
} \\
J_{5}(\frac{\partial }{\partial x_{5n+i}})=-\frac{\partial }{\partial x_{i}}
\\
J_{5}(\frac{\partial }{\partial x_{6n+i}})=-\frac{\partial }{\partial
x_{4n+i}} \\
J_{5}(\frac{\partial }{\partial x_{7n+i}})=\frac{\partial }{\partial x_{2n+i}%
}%
\end{array}
&
\begin{array}{c}
J_{6}(\frac{\partial }{\partial x_{i}})=\frac{\partial }{\partial x_{6n+i}}
\\
J_{6}(\frac{\partial }{\partial x_{n+i}})=-\frac{\partial }{\partial x_{7n+i}%
} \\
J_{6}(\frac{\partial }{\partial x_{2n+i}})=-\frac{\partial }{\partial
x_{3n+i}} \\
J_{6}(\frac{\partial }{\partial x_{3n+i}})=\frac{\partial }{\partial x_{2n+i}%
} \\
J_{6}(\frac{\partial }{\partial x_{4n+i}})=\frac{\partial }{\partial x_{5n+i}%
} \\
J_{6}(\frac{\partial }{\partial x_{5n+i}})=-\frac{\partial }{\partial
x_{4n+i}} \\
\text{ }J_{6}(\frac{\partial }{\partial x_{6n+i}})=-\frac{\partial }{%
\partial x_{i}} \\
J_{6}(\frac{\partial }{\partial x_{7n+i}})=\frac{\partial }{\partial x_{n+i}}%
.%
\end{array}%
\end{array}
\label{2.5}
\end{equation}

A canonical local basis$\{J_{1}^{\ast },J_{2}^{\ast },J_{3}^{\ast
},J_{4}^{\ast },J_{5}^{\ast },J_{6}^{\ast }\}$ of $V^{\ast }$ of the
cotangent space $T^{\ast }(M)$ of manifold $M$ satisfies the following
condition:

\begin{equation}
J_{1}^{\ast 2}=J_{2}^{\ast 2}=\text{ }J_{3}^{\ast 2}=J_{4}^{\ast
2}=J_{5}^{\ast 2}=\text{ }J_{6}^{\ast 2}=-I,  \label{2.6}
\end{equation}%
being%
\begin{equation}
\begin{array}{ccc}
\begin{array}{c}
J_{1}^{\ast }(dx_{i})=dx_{n+i} \\
\text{ }J_{1}^{\ast }(dx_{n+i})=-dx_{i} \\
J_{1}^{\ast }(dx_{2n+i})=dx_{4n+i} \\
\text{ }J_{1}^{\ast }(dx_{3n+i})=dx_{5n+i} \\
J_{1}^{\ast }(dx_{4n+i})=-dx_{2n+i} \\
\text{ }J_{1}^{\ast }(dx_{5n+i})=-dx_{3n+i} \\
\text{ }J_{1}^{\ast }(dx_{6n+i})=dx_{7n+i} \\
\text{ }J_{1}^{\ast }(dx_{7n+i})=-dx_{6n+i}%
\end{array}
&
\begin{array}{c}
J_{2}^{\ast }(dx_{i})=dx_{2n+i} \\
J_{2}^{\ast }(dx_{n+i})=-dx_{4n+i} \\
\text{ }J_{2}^{\ast }(dx_{2n+i})=-dx_{i} \\
J_{2}^{\ast }(dx_{3n+i})=dx_{6n+i} \\
J_{2}^{\ast }(dx_{4n+i})=dx_{n+i} \\
J_{2}^{\ast }(dx_{5n+i})=-dx_{7n+i} \\
\text{ }J_{2}^{\ast }(dx_{6n+i})=-dx_{3n+i} \\
J_{2}^{\ast }(dx_{7n+i})=dx_{5n+i}%
\end{array}
&
\begin{array}{c}
J_{3}^{\ast }(dx_{i})=dx_{3n+i} \\
\text{ }J_{3}^{\ast }(dx_{n+i})=-dx_{5n+i} \\
J_{3}^{\ast }(dx_{2n+i})=-dx_{6n+i} \\
J_{3}^{\ast }(dx_{3n+i})=-dx_{i} \\
J_{3}^{\ast }(dx_{4n+i})=dx_{7n+i} \\
\text{ }J_{3}^{\ast }(dx_{5n+i})=dx_{n+i} \\
\text{ }J_{3}^{\ast }(dx_{6n+i})=dx_{2n+i} \\
\text{ }J_{3}^{\ast }(dx_{7n+i})=-dx_{4n+i}%
\end{array}
\\
\begin{array}{c}
\text{ }J_{4}^{\ast }(dx_{i})=dx_{4n+i} \\
J_{4}^{\ast }(dx_{n+i})=-dx_{2n+i} \\
J_{4}^{\ast }(dx_{2n+i})=dx_{n+i} \\
\text{ }J_{4}^{\ast }(dx_{3n+i})=-dx_{7n+i} \\
J_{4}^{\ast }(dx_{4n+i})=-dx_{i} \\
J_{4}^{\ast }(dx_{5n+i})=dx_{6n+i} \\
\text{ }J_{4}^{\ast }(dx_{6n+i})=-dx_{5n+i} \\
J_{4}^{\ast }(dx_{7n+i})=dx_{3n+i}%
\end{array}
&
\begin{array}{c}
J_{5}^{\ast }(dx_{i})=dx_{5n+i} \\
J_{5}^{\ast }(dx_{n+i})=-dx_{3n+i} \\
\text{ }J_{5}^{\ast }(dx_{2n+i})=-dx_{7n+i} \\
J_{5}^{\ast }(dx_{3n+i})=dx_{n+i} \\
J_{5}^{\ast }(dx_{4n+i})=dx_{6n+i} \\
J_{5}^{\ast }(dx_{5n+i})=-dx_{i} \\
\text{ }J_{5}^{\ast }(dx_{6n+i})=-dx_{4n+i} \\
\text{ }J_{5}^{\ast }(dx_{7n+i})=dx_{2n+i}%
\end{array}
&
\begin{array}{c}
J_{6}^{\ast }(dx_{i})=dx_{6n+i} \\
\text{ }J_{6}^{\ast }(dx_{n+i})=-dx_{7n+i} \\
\text{ }J_{6}^{\ast }(dx_{2n+i})=-dx_{3n+i} \\
\text{ }J_{6}^{\ast }(dx_{3n+i})=dx_{2n+i} \\
J_{6}^{\ast }(dx_{4n+i})=dx_{5n+i} \\
\text{ }J_{6}^{\ast }(dx_{5n+i})=-dx_{4n+i} \\
\text{ }J_{6}^{\ast }(dx_{6n+i})=-dx_{i} \\
\text{ }J_{6}^{\ast }(dx_{7n+i})=dx_{n+i}.%
\end{array}%
\end{array}
\label{2.7}
\end{equation}

\section{Hamilton Mechanics}

In this section, we obtain Hamilton equations and Hamilton mechanical system
for quantum and classical mechanics by means of a canonical local basis $%
\{J_{1}^{\ast },J_{2}^{\ast },J_{3}^{\ast },J_{4}^{\ast },J_{5}^{\ast
},J_{6}^{\ast }\}$ of $V$ on Clifford K\"{a}hler manifold $(M,V).$ We saw
that the Hamilton equations using basis $\{J_{1}^{\ast },J_{2}^{\ast
},J_{3}^{\ast }\}$ of $V$ on $(\mathbf{R}^{8n},V)$ are introduced in \cite%
{tekkoyun2}$.$ In this study, it is seen that they are the same as the
equations \ obtained by operators $J_{1}^{\ast },J_{2}^{\ast },J_{3}^{\ast }$
of $V$ on Clifford K\"{a}hler manifold $(M,V).$ If we redetermine them, they
are respectively:

first:%
\begin{equation*}
\begin{array}{c}
\frac{dx_{i}}{dt}=-\frac{\partial \mathbf{H}}{\partial x_{n+i}},\text{ }%
\frac{dx_{n+i}}{dt}=\frac{\partial \mathbf{H}}{\partial x_{i}},\text{ }\frac{%
dx_{2n+i}}{dt}=-\frac{\partial \mathbf{H}}{\partial x_{4n+i}},\text{ }\frac{%
dx_{3n+i}}{dt}=-\frac{\partial \mathbf{H}}{\partial x_{5n+i}}, \\
\frac{dx_{4n+i}}{dt}=\frac{\partial \mathbf{H}}{\partial x_{2n+i}},\text{ }%
\frac{dx_{5n+i}}{dt}=\frac{\partial \mathbf{H}}{\partial x_{3n+i}},\text{ }%
\frac{dx_{6n+i}}{dt}=-\frac{\partial \mathbf{H}}{\partial x_{7n+i}},\text{ }%
\frac{dx_{7n+i}}{dt}=\frac{\partial \mathbf{H}}{\partial x_{6n+i}}.%
\end{array}%
\end{equation*}

second:

\begin{equation*}
\begin{array}{c}
\frac{dx_{i}}{dt}=-\frac{\partial \mathbf{H}}{\partial x_{2n+i}},\text{ }%
\frac{dx_{n+i}}{dt}=\frac{\partial \mathbf{H}}{\partial x_{4n+i}},\text{ }%
\frac{dx_{2n+i}}{dt}=\frac{\partial \mathbf{H}}{\partial x_{i}},\text{ }%
\frac{dx_{3n+i}}{dt}=-\frac{\partial \mathbf{H}}{\partial x_{6n+i}}, \\
\frac{dx_{4n+i}}{dt}=-\frac{\partial \mathbf{H}}{\partial x_{n+i}},\text{ }%
\frac{dx_{5n+i}}{dt}=\frac{\partial \mathbf{H}}{\partial x_{7n+i}},\text{ }%
\frac{dx_{6n+i}}{dt}=\frac{\partial \mathbf{H}}{\partial x_{3n+i}},\text{ }%
\frac{dx_{7n+i}}{dt}=-\frac{\partial \mathbf{H}}{\partial x_{5n+i}}.%
\end{array}%
\end{equation*}

third:

\begin{equation*}
\begin{array}{c}
\frac{dx_{i}}{dt}=-\frac{\partial \mathbf{H}}{\partial x_{3n+i}},\text{ }%
\frac{dx_{n+i}}{dt}=\frac{\partial \mathbf{H}}{\partial x_{5n+i}},\text{ }%
\frac{dx_{2n+i}}{dt}=\frac{\partial \mathbf{H}}{\partial x_{6n+i}},\text{ }%
\frac{dx_{3n+i}}{dt}=\frac{\partial \mathbf{H}}{\partial x_{i}}, \\
\frac{dx_{4n+i}}{dt}=-\frac{\partial \mathbf{H}}{\partial x_{7n+i}},\text{ }%
\frac{dx_{5n+i}}{dt}=-\frac{\partial \mathbf{H}}{\partial x_{n+i}},\text{ }%
\frac{dx_{6n+i}}{dt}=-\frac{\partial \mathbf{H}}{\partial x_{2n+i}},\text{ }%
\frac{dx_{7n+i}}{dt}=\frac{\partial \mathbf{H}}{\partial x_{4n+i}}.%
\end{array}%
\end{equation*}

Fourth, let $(M,V)$ be a Clifford K\"{a}hler manifold. Suppose that a
component of almost Clifford structure $V^{\ast }$, a Liouville form and a
1-form on Clifford K\"{a}hler manifold $(M,V)$ are given by $J_{4}^{\ast }$,
$\lambda _{J_{4}^{\ast }}$ and $\omega _{J_{4}^{\ast }}$, respectively$.$

Putting
\begin{eqnarray*}
\omega _{J_{4}^{\ast }} &=&\frac{1}{2}%
(x_{i}dx_{i}+x_{n+i}dx_{n+i}+x_{2n+i}dx_{2n+i}+x_{3n+i}dx_{3n+i} \\
&&+x_{4n+i}dx_{4n+i}+x_{5n+i}dx_{5n+i}+x_{6n+i}dx_{6n+i}+x_{7n+i}dx_{7n+i}),
\end{eqnarray*}%
we have%
\begin{eqnarray*}
\lambda _{J_{4}^{\ast }} &=&J_{4}^{\ast }(\omega _{J_{4}^{\ast }})=\frac{1}{2%
}(x_{i}dx_{4n+i}-x_{n+i}dx_{2n+i}+x_{2n+i}dx_{n+i}-x_{3n+i}dx_{7n+i} \\
&&-x_{4n+i}dx_{i}+x_{5n+i}dx_{6n+i}-x_{6n+i}dx_{5n+i}+x_{7n+i}dx_{3n+i}).
\end{eqnarray*}%
It is known that if $\Phi _{J_{4}^{\ast }}$ is a closed K\"{a}hler form on
Clifford K\"{a}hler manifold $(M,V),$ then $\Phi _{J_{4}^{\ast }}$ is also a
symplectic structure on Clifford K\"{a}hler manifold $(M,V)$.

Take into consideration that Hamilton vector field $X$ associated with
Hamilton energy $\mathbf{H}$ is given by%
\begin{equation}
\begin{array}{c}
X=X^{i}\frac{\partial }{\partial x_{i}}+X^{n+i}\frac{\partial }{\partial
x_{n+i}}+X^{2n+i}\frac{\partial }{\partial x_{2n+i}}+X^{3n+i}\frac{\partial
}{\partial x_{3n+i}} \\
+X^{4n+i}\frac{\partial }{\partial x_{4n+i}}+X^{5n+i}\frac{\partial }{%
\partial x_{5n+i}}+X^{6n+i}\frac{\partial }{\partial x_{6n+i}}+X^{7n+i}\frac{%
\partial }{\partial x_{7n+i}}.%
\end{array}
\label{4.2}
\end{equation}

Then
\begin{equation}
\Phi _{J_{4}^{\ast }}=-d\lambda _{J_{4}^{\ast }}=dx_{n+i}\wedge
dx_{2n+i}+dx_{3n+i}\wedge dx_{7n+i}+dx_{4n+i}\wedge dx_{i}+dx_{6n+i}\wedge
dx_{5n+i}  \label{4.3}
\end{equation}%
and%
\begin{equation}
\begin{array}{c}
i_{X}\Phi _{J_{4}^{\ast }}=\Phi _{J_{4}^{\ast
}}(X)=X^{n+i}dx_{2n+i}-X^{2n+i}dx_{n+i}+X^{3n+i}dx_{7n+i}-X^{7n+i}dx_{3n+i}
\\
+X^{4n+i}dx_{i}-X^{i}dx_{4n+i}+X^{6n+i}dx_{5n+i}-X^{5n+i}dx_{6n+i}.%
\end{array}
\label{4.4}
\end{equation}%
Furthermore, the differential of Hamilton energy is obtained as follows:%
\begin{equation}
\begin{array}{c}
d\mathbf{H}=\frac{\partial \mathbf{H}}{\partial x_{i}}dx_{i}+\frac{\partial
\mathbf{H}}{\partial x_{n+i}}dx_{n+i}+\frac{\partial \mathbf{H}}{\partial
x_{2n+i}}dx_{2n+i}+\frac{\partial \mathbf{H}}{\partial x_{3n+i}}dx_{3n+i} \\
+\frac{\partial \mathbf{H}}{\partial x_{4n+i}}dx_{4n+i}+\frac{\partial
\mathbf{H}}{\partial x_{5n+i}}dx_{5n+i}+\frac{\partial \mathbf{H}}{\partial
x_{6n+i}}dx_{6n+i}+\frac{\partial \mathbf{H}}{\partial x_{7n+i}}dx_{7n+i}.%
\end{array}
\label{4.5}
\end{equation}%
According to \textbf{Eq.}(\ref{1.1}), if equaled \textbf{Eq. }(\ref{4.4})
and \textbf{Eq. }(\ref{4.5}), the Hamilton vector field is calculated as
follows:%
\begin{equation}
\begin{array}{c}
X=-\frac{\partial \mathbf{H}}{\partial x_{4n+i}}\frac{\partial }{\partial
x_{i}}+\frac{\partial \mathbf{H}}{\partial x_{2n+i}}\frac{\partial }{%
\partial x_{n+i}}-\frac{\partial \mathbf{H}}{\partial x_{n+i}}\frac{\partial
}{\partial x_{2n+i}}+\frac{\partial \mathbf{H}}{\partial x_{7n+i}}\frac{%
\partial }{\partial x_{3n+i}} \\
+\frac{\partial \mathbf{H}}{\partial x_{i}}\frac{\partial }{\partial x_{4n+i}%
}-\frac{\partial \mathbf{H}}{\partial x_{6n+i}}\frac{\partial }{\partial
x_{5n+i}}+\frac{\partial \mathbf{H}}{\partial x_{5n+i}}\frac{\partial }{%
\partial x_{6n+i}}-\frac{\partial \mathbf{H}}{\partial x_{3n+i}}\frac{%
\partial }{\partial x_{7n+i}}%
\end{array}
\label{4.6}
\end{equation}

Assume that a curve
\begin{equation}
\alpha :\mathbf{R}\rightarrow M  \label{4.7}
\end{equation}%
be an integral curve of the Hamilton vector field $X$, i.e.,
\begin{equation}
X(\alpha (t))=\overset{.}{\alpha },\,\,t\in \mathbf{R}.  \label{4.8}
\end{equation}%
In the local coordinates, it is found that
\begin{equation}
\alpha
(t)=(x_{i},x_{n+i},x_{2n+i},x_{3n+i},x_{4n+i},x_{5n+i},x_{6n+i},x_{7n+i})
\label{4.9}
\end{equation}%
and%
\begin{equation}
\begin{array}{c}
\overset{.}{\alpha }(t)=\frac{dx_{i}}{dt}\frac{\partial }{\partial x_{i}}+%
\frac{dx_{n+i}}{dt}\frac{\partial }{\partial x_{n+i}}+\frac{dx_{2n+i}}{dt}%
\frac{\partial }{\partial x_{2n+i}}+\frac{dx_{3n+i}}{dt}\frac{\partial }{%
\partial x_{3n+i}} \\
+\frac{dx_{4n+i}}{dt}\frac{\partial }{\partial x_{4n+i}}+\frac{dx_{5n+i}}{dt}%
\frac{\partial }{\partial x_{5n+i}}+\frac{dx_{6n+i}}{dt}\frac{\partial }{%
\partial x_{6n+i}}+\frac{dx_{7n+i}}{dt}\frac{\partial }{\partial x_{7n+i}}.%
\end{array}
\label{4.10}
\end{equation}%
Thinking out \textbf{Eq. }(\ref{4.8}), if equaled \textbf{Eq. }(\ref{4.6})
and\textbf{\ Eq. }(\ref{4.10}), it follows%
\begin{equation}
\begin{array}{c}
\frac{dx_{i}}{dt}=-\frac{\partial \mathbf{H}}{\partial x_{4n+i}},\text{ }%
\frac{dx_{n+i}}{dt}=\frac{\partial \mathbf{H}}{\partial x_{2n+i}},\text{ }%
\frac{dx_{2n+i}}{dt}=-\frac{\partial \mathbf{H}}{\partial x_{n+i}},\text{ }%
\frac{dx_{3n+i}}{dt}=\frac{\partial \mathbf{H}}{\partial x_{7n+i}}, \\
\frac{dx_{4n+i}}{dt}=\frac{\partial \mathbf{H}}{\partial x_{i}},\text{ }%
\frac{dx_{5n+i}}{dt}=-\frac{\partial \mathbf{H}}{\partial x_{6n+i}},\text{ }%
\frac{dx_{6n+i}}{dt}=\frac{\partial \mathbf{H}}{\partial x_{5n+i}},\text{ }%
\frac{dx_{7n+i}}{dt}=-\frac{\partial \mathbf{H}}{\partial x_{3n+i}}.%
\end{array}
\label{4.11}
\end{equation}%
Hence, the equations obtained in \textbf{Eq. }(\ref{4.11}) are shown to be
\textit{Hamilton equations} with respect to component $J_{4}^{\ast }$ of
almost Clifford structure $V^{\ast }$ on Clifford K\"{a}hler manifold $%
(M,V), $ and then the triple $(M,,\Phi _{J_{4}^{\ast }},X)$ is said to be a
\textit{Hamilton mechanical system }on Clifford K\"{a}hler manifold $(M,V)$.

Fifth, let $(M,V)$ be a Clifford K\"{a}hler manifold. Assume that an element
of almost Clifford structure $V^{\ast }$, a Liouville form and a 1-form on
Clifford K\"{a}hler manifold $(M,V)$ are determined by $J_{5}^{\ast }$, $%
\lambda _{J_{5}^{\ast }}$ and $\omega _{J_{5}^{\ast }}$, respectively$.$

Setting
\begin{eqnarray*}
\omega _{J_{5}^{\ast }} &=&\frac{1}{2}%
(x_{i}dx_{i}+x_{n+i}dx_{n+i}+x_{2n+i}dx_{2n+i}+x_{3n+i}dx_{3n+i} \\
&&+x_{4n+i}dx_{4n+i}+x_{5n+i}dx_{5n+i}+x_{6n+i}dx_{6n+i}+x_{7n+i}dx_{7n+i}),
\end{eqnarray*}%
we have
\begin{eqnarray*}
\lambda _{J_{5}^{\ast }} &=&J_{5}^{\ast }(\omega _{J_{5}^{\ast }})=\frac{1}{2%
}(x_{i}dx_{5n+i}-x_{n+i}dx_{3n+i}-x_{2n+i}dx_{7n+i}+x_{3n+i}dx_{n+i} \\
&&+x_{4n+i}dx_{6n+i}-x_{5n+i}dx_{i}-x_{6n+i}dx_{4n+i}+x_{7n+i}dx_{2n+i}).
\end{eqnarray*}

Assume that $X$ is a Hamilton vector field related to Hamilton energy $%
\mathbf{H}$ and given by \textbf{Eq. }(\ref{4.2}).

Take into consideration
\begin{equation}
\Phi _{J_{5}^{\ast }}=-d\lambda _{J_{5}^{\ast }}=dx_{n+i}\wedge
dx_{3n+i}+dx_{2n+i}\wedge dx_{7n+i}+dx_{5n+i}\wedge dx_{i}+dx_{6n+i}\wedge
dx_{4n+i},  \label{4.12}
\end{equation}%
then we find%
\begin{equation}
\begin{array}{c}
i_{X}\Phi _{J_{5}^{\ast }}=\Phi _{J_{5}^{\ast
}}(X)=X^{n+i}dx_{3n+i}-X^{3n+i}dx_{n+i}+X^{2n+i}dx_{7n+i}-X^{7n+i}dx_{2n+i}
\\
+X^{5n+i}dx_{i}-X^{i}dx_{5n+i}+X^{6n+i}dx_{4n+i}-X^{4n+i}dx_{6n+i}.%
\end{array}
\label{4.13}
\end{equation}%
According to \textbf{Eq.}(\ref{1.1}), if we equal \textbf{Eq. }(\ref{4.5})
and \textbf{Eq. }(\ref{4.13}), it follows%
\begin{equation}
\begin{array}{c}
X=-\frac{\partial \mathbf{H}}{\partial x_{5n+i}}\frac{\partial }{\partial
x_{i}}+\frac{\partial \mathbf{H}}{\partial x_{3n+i}}\frac{\partial }{%
\partial x_{n+i}}+\frac{\partial \mathbf{H}}{\partial x_{7n+i}}\frac{%
\partial }{\partial x_{2n+i}}-\frac{\partial \mathbf{H}}{\partial x_{n+i}}%
\frac{\partial }{\partial x_{3n+i}} \\
-\frac{\partial \mathbf{H}}{\partial x_{6n+i}}\frac{\partial }{\partial
x_{4n+i}}+\frac{\partial \mathbf{H}}{\partial x_{i}}\frac{\partial }{%
\partial x_{5n+i}}+\frac{\partial \mathbf{H}}{\partial x_{4n+i}}\frac{%
\partial }{\partial x_{6n+i}}-\frac{\partial \mathbf{H}}{\partial x_{2n+i}}%
\frac{\partial }{\partial x_{7n+i}}%
\end{array}
\label{4.14}
\end{equation}

Taking \textbf{Eq. }(\ref{4.8}), \textbf{Eqs. }(\ref{4.10}) and\textbf{\ }(%
\ref{4.14}) are equal, we obtain equations%
\begin{equation}
\begin{array}{c}
\frac{dx_{i}}{dt}=-\frac{\partial \mathbf{H}}{\partial x_{5n+i}},\text{ }%
\frac{dx_{n+i}}{dt}=\frac{\partial \mathbf{H}}{\partial x_{3n+i}},\text{ }%
\frac{dx_{2n+i}}{dt}=\frac{\partial \mathbf{H}}{\partial x_{7n+i}},\text{ }%
\frac{dx_{3n+i}}{dt}=-\frac{\partial \mathbf{H}}{\partial x_{n+i}}, \\
\frac{dx_{4n+i}}{dt}=-\frac{\partial \mathbf{H}}{\partial x_{6n+i}},\text{ }%
\frac{dx_{5n+i}}{dt}=\frac{\partial \mathbf{H}}{\partial x_{i}},\text{ }%
\frac{dx_{6n+i}}{dt}=\frac{\partial \mathbf{H}}{\partial x_{4n+i}},\text{ }%
\frac{dx_{7n+i}}{dt}=-\frac{\partial \mathbf{H}}{\partial x_{2n+i}}.%
\end{array}
\label{4.15}
\end{equation}%
In the end, the equations found in \textbf{Eq. }(\ref{4.15}) are seen to be
\textit{Hamilton equations} with respect to component $J_{5}^{\ast }$ of \
almost Clifford structure $V^{\ast }$ on Clifford K\"{a}hler manifold $%
(M,V), $ and then the triple $(M,\Phi _{J_{5}^{\ast }},X)$ is named a
\textit{Hamilton mechanical system }on Clifford K\"{a}hler manifold $(M,V)$.

Sixth, let $(M,V)$ be a Clifford K\"{a}hler manifold. By $J_{6}^{\ast }$ $%
\lambda _{J_{6}^{\ast }}$ and $\omega _{J_{6}^{\ast }},$ we denote a
component of almost Clifford structure $V^{\ast }$, a Liouville form and a
1-form on Clifford K\"{a}hler manifold $(M,V)$, respectively$.$

Let $\omega _{J_{6}^{\ast }}$ be determined by

\begin{eqnarray*}
\omega _{J_{6}^{\ast }} &=&\frac{1}{2}%
(x_{i}dx_{i}+x_{n+i}dx_{n+i}+x_{2n+i}dx_{2n+i}+x_{3n+i}dx_{3n+i} \\
&&+x_{4n+i}dx_{4n+i}+x_{5n+i}dx_{5n+i}+x_{6n+i}dx_{6n+i}+x_{7n+i}dx_{7n+i})
\end{eqnarray*}%
Then it yields

\begin{eqnarray*}
\lambda _{J_{6}^{\ast }} &=&J_{6}^{\ast }(\omega _{J_{6}^{\ast }})=\frac{1}{2%
}(x_{i}dx_{6n+i}-x_{n+i}dx_{7n+i}-x_{2n+i}dx_{3n+i}+x_{3n+i}dx_{2n+i} \\
&&+x_{4n+i}dx_{5n+i}-x_{5n+i}dx_{4n+i}-x_{6n+i}dx_{i}+x_{7n+i}dx_{n+i}).
\end{eqnarray*}%
It is known that if $\Phi _{J_{6}^{\ast }}$ is a closed K\"{a}hler form on
Clifford K\"{a}hler manifold $(M,V),$ then $\Phi _{J_{6}^{\ast }}$ is also a
symplectic structure on Clifford K\"{a}hler manifold $(M,V)$.

Take $X$ . It is Hamilton vector field connected with Hamilton energy $%
\mathbf{H}$ and given by \textbf{Eq. }(\ref{4.2}).

Considering
\begin{equation}
\Phi _{J_{6}^{\ast }}=-d\lambda _{J_{6}^{\ast }}=dx_{n+i}\wedge
dx_{7n+i}+dx_{2n+i}\wedge dx_{3n+i}+dx_{5n+i}\wedge
dx_{4n+i}+dx_{6n+i}\wedge dx_{i},  \label{4.17}
\end{equation}%
we calculate%
\begin{equation}
\begin{array}{c}
i_{X}\Phi _{J_{6}^{\ast }}=\Phi _{J_{6}^{\ast
}}(X)=X^{n+i}dx_{7n+i}-X^{7n+i}dx_{n+i}+X^{2n+i}dx_{3n+i}-X^{3n+i}dx_{2n+i}
\\
+X^{5n+i}dx_{4n+i}-X^{4n+i}dx_{5n+i}+X^{6n+i}dx_{i}-X^{i}dx_{6n+i}.%
\end{array}
\label{4.18}
\end{equation}%
According to \textbf{Eq.}(\ref{1.1}), \textbf{Eqs. }(\ref{4.5}) and (\ref%
{4.18}) are equaled, Hamilton vector field is found as follows:%
\begin{equation}
\begin{array}{c}
X=-\frac{\partial \mathbf{H}}{\partial x_{6n+i}}\frac{\partial }{\partial
x_{i}}+\frac{\partial \mathbf{H}}{\partial x_{7n+i}}\frac{\partial }{%
\partial x_{n+i}}+\frac{\partial \mathbf{H}}{\partial x_{3n+i}}\frac{%
\partial }{\partial x_{2n+i}}-\frac{\partial \mathbf{H}}{\partial x_{2n+i}}%
\frac{\partial }{\partial x_{3n+i}} \\
-\frac{\partial \mathbf{H}}{\partial x_{5n+i}}\frac{\partial }{\partial
x_{4n+i}}+\frac{\partial \mathbf{H}}{\partial x_{4n+i}}\frac{\partial }{%
\partial x_{5n+i}}+\frac{\partial \mathbf{H}}{\partial x_{i}}\frac{\partial
}{\partial x_{6n+i}}-\frac{\partial \mathbf{H}}{\partial x_{n+i}}\frac{%
\partial }{\partial x_{7n+i}}.%
\end{array}
\label{4.19}
\end{equation}

Considering \textbf{Eq. }(\ref{4.8}), we equal \textbf{Eq. }(\ref{4.10}) and%
\textbf{\ Eq. }(\ref{4.19}), it holds%
\begin{equation}
\begin{array}{c}
\frac{dx_{i}}{dt}=-\frac{\partial \mathbf{H}}{\partial x_{6n+i}},\text{ }%
\frac{dx_{n+i}}{dt}=\frac{\partial \mathbf{H}}{\partial x_{7n+i}},\text{ }%
\frac{dx_{2n+i}}{dt}=\frac{\partial \mathbf{H}}{\partial x_{3n+i}},\text{ }%
\frac{dx_{3n+i}}{dt}=-\frac{\partial \mathbf{H}}{\partial x_{2n+i}}, \\
\frac{dx_{4n+i}}{dt}=-\frac{\partial \mathbf{H}}{\partial x_{5n+i}},\text{ }%
\frac{dx_{5n+i}}{dt}=\frac{\partial \mathbf{H}}{\partial x_{4n+i}},\text{ }%
\frac{dx_{6n+i}}{dt}=\frac{\partial \mathbf{H}}{\partial x_{i}},\text{ }%
\frac{dx_{7n+i}}{dt}=-\frac{\partial \mathbf{H}}{\partial x_{n+i}}.%
\end{array}
\label{4.20}
\end{equation}%
Finally, the equations calculated in \textbf{Eq. }(\ref{4.20}) are called to
be \textit{Hamilton equations} with respect to component $J_{6}^{\ast }$ of
almost Clifford structure $V^{\ast }$ on Clifford K\"{a}hler manifold $%
(M,V), $ and then the triple $(M,\Phi _{J_{6}^{\ast }},X)$ is said to be a
\textit{Hamilton mechanical system }on Clifford K\"{a}hler manifold $(M,V)$.

\section{Conclusion}

Hamilton Formalisms has intrinsically been described with taking into
account the basis $\{J_{1}^{\ast },J_{2}^{\ast },J_{3}^{\ast },J_{4}^{\ast
},J_{5}^{\ast },J_{6}^{\ast }\}$ of almost Clifford structure $V^{\ast }$ on
Clifford K\"{a}hler manifold $(M,V)$.

Hamilton models arise to be a very important tool since they present a
simple method to describe the model for dynamical systems. In solving
problems in classical mechanics, the rotational mechanical system will then
be easily usable model.

Since a new model for dynamic systems on subspaces and spaces is needed,
equations (\ref{4.11}), (\ref{4.15}) and (\ref{4.20}) are only considered to
be a first step to realize how Clifford geometry has been used in
understanding, modeling \ and solving problems in different physical fields.

For further research, the Hamilton vector fields and equations obtained here
are advised to deal with problems in applicable fields of quantum and
classical mechanics of physics.

\end{document}